\documentclass[twocolumn,showpacs,showkeywords,prb]{revtex4}
\usepackage{graphicx}
\usepackage{dcolumn}
\usepackage{amsmath}
\usepackage{latexsym}

\begin{document}
\title
{Resonant-bar detectors of gravitational wave as possible probe of the noncommutative structure of space}
\author{Anirban Saha}
\email{anirban@iucaa.ernet.in }
\altaffiliation{Visiting Associate in Inter University Centre for Astronomy
and Astrophysics, Pune, India}
\author{Sunandan Gangopadhyay}
\email{sunandan.gangopadhyay@gmail.com , sunandan@iucaa.ernet.in}
\altaffiliation{Visiting Associate in Inter University Centre for Astronomy $\&$ Astrophysics (IUCAA), Pune, India}
\affiliation{Department of Physics, West Bengal State University,Barasat, North 24 Paraganas, West Bengal, India}

\begin{abstract}
{\noindent 
We report the plausibility of using quantum mechanical transitions, induced by the combined effect of Gravitational wave (GW) and noncommutative (NC) structure of space, among the states of a 2-dimensional harmonic oscillator, to probe the spatial NC geometry. The phonon modes excited by the passing GW within the resonant bar-detector are formally identical to forced harmonic oscillator and they represent a length variation of roughly the same order of magnitude as the characteristic length-scale of spatial noncommutativity estimated from the phenomenological upper bound of the NC parameter. This motivates our present work.
We employ various GW wave-forms that are typically expected from possible astronomical sources. We find that the transition probablities are quite sensitive to the nature of polarization of the GW. We also elaborate on the particular type of sources of GW, radiation from which can induce transitions that can be used as effective probe of the spatial noncommutative structure.
}
\end{abstract}


\maketitle

Gravitational waves (GW(s)) are small ripples in the fabric of spacetime. The present day GW detectors primarily consists of ground-based (LIGO, VIRGO, GEO, TAMA etc.)  and space-based (LISA) interferometers \cite {GW-detection_status}. 
However, the search of GWs began with resonant-mass detectors, pioneered by Weber in the 60's \cite{Weber_1, Weber_2, Weber_3}. In the decades that followed, the sensitivity of resonant-mass detectors have improved considerably, though it is clear that it could only detect relatively strong signals from within our Galaxy or the immediate galactic neighbourhood \cite{bar_1}. 
Nevertheless,
the study of resonant-bar detectors is fundamental because it focuses on how GW interacts with elastic matter causing vibrations with amplitudes many order smaller than the size of a nucleus. In a bar detector it is possible to detect these tiny vibrations corresponding to just a few tens of phonons \cite{Magg}, and variations $\Delta L$ of the bar-length $L \sim 1 {\rm m}$, with $\frac{\Delta L}{L} \sim 10^{-19}$. 
 
Interestingly, it has long been suggested in various Gedenken experiments that a sharp localization of events in space would induce an uncertainty in spatial coordinates \cite{Dop, Alu} at the quantum level. This uncertainty can be realized by imposing the NC Heisenberg algebra on the operators representing phase-space variables 
\begin{eqnarray}
\left[{\hat x}_{i}, {\hat p}_{j}\right] = i\hbar \delta_{ij} \>, \quad 
\left[{\hat x}_{i}, {\hat x}_{j}\right] = i \theta_{ij}= i \theta \epsilon_{ij} \>,\quad 
\left[{\hat p}_{i}, {\hat p}_{j}\right] = 0\>.
\label{e9a}
\end{eqnarray}
where $\theta_{ij}$ is the constant antisymmetric tensor, which is writen in terms of the constant NC parameter $\theta$ and the totally antisymmetric  tensor $\epsilon_{ij}$. Such granularity in spatial structure have been motivated by string theoretic \cite{SW} and quantum gravity \cite{hogan} results also.

A wide range of theories, dubbed the NC theories, have been constructed in this framework. This includs NC quantum mechanics (NCQM) \cite{duv, nair, duv1, mezin, rban, anisaha, hazra, gov, sgprl}, NC quantum field theory (NCQFT) \cite{doug, masud, bal}, gauge theories \cite{szabo} and gravity \cite{grav, grav1, banerjee11, sgrb}. Certain possible phenomenological consequences \cite{jabbari1, rs5, rs7, rs9, rs10, sun} have also been predicted. Naturally, a part of the endeavour is spent in finding the order of the NC parameter and exploring its connection with observations \cite{mpr, cst, carol, bert0, RB, ani, stern}. The upperbound on the coordinate commutator $|\theta|$ found in \cite{carol} is $\lesssim \left(10 {\rm TeV}\right)^{-2}$ which corresponds to $4 \times 10^{-40} {\rm m}^{2}$ for $\hbar$$=$$c$$=$$1$.\footnote{In a more general NC space-time structure \cite{SW} given by $\left[x^{\mu}, x^{\nu}\right] = i \theta^{\mu \nu}$ such upperbounds on time-space NC parameter is\cite{ani} $\theta^{0i}$ $\lesssim 9.51\times 10^{-18} {\rm m}^{2}$.} However, recent studies in NCQM revealed that the NC parameter associated with different particles may not be the same \cite{pmho, vassilavich} and this bound could be as high as $\theta \lesssim \left(4 {\rm GeV}\right)^{-2} - \left(30 {\rm MeV}\right)^{-2}$  \cite{stern}. These upperbounds correspond to the length scale $\sim 10^{-20} {\rm m} - 10^{-17} {\rm m}$ which overlaps the length scale where the first GW detection is expected. Thus a good possibility of detecting the NC structure of spacetime would be in the GW detection experiments as it may as well pick up the NC signature of space-time. For this purpose we need NCQM of GW-matter interaction that can anticiapte the NC effects in GW detection events.

With this motivation, we have studied the interaction of GW(s) with simple matter systems in a NCQM framework in \cite{ncgw1, ncgw2, ncgw_3, ncgw_4}. Our particular interest is in the NCQM of harmonic oscillator (HO) interacting with GW because the response of a bar-detector to GW can be cast as phonon mode excitations formally identical to forced HO \cite{Magg}.
In \cite{ncgw_4} using system parameters typical to the resonant bar detectors and existing upper-bound of the spatial NC parameter we have shown that our NCQM analysis is likely to be relevant for identifying NC effects in the detector read-outs. Thus, NCQM of the HO interacting with GW is of fundamental importance. Therefore, we investigate the transition probabilities between the ground state and the excited states of this system in the present paper by treating the combined effect of GW and spatial noncommutativity as time-dependent perturbations. We employ a number of different GW wave-forms that are typically expected from runaway astronomical events.

To proceed we first obtain the classical Hamiltonian appropriate for the GW-HO interaction system. This can be simply done by noting that in the proper detector frame the geodesic deviation equation for a $2-$dimensional harmonic oscillator of mass $m$  and frequency $\varpi$ subject to linearized GW becomes \cite{Magg} 
\begin{equation}
m \ddot{{x}} ^{j}= - m{R^j}_{0,k0} {x}^{k} - m \varpi^{2} x^{j}
\label{e5}
\end{equation}
where dot denotes derivative with respect to the coordinate time of the proper detector frame\footnote{It is the same as it's proper time to first order in the metric perturbation.}, ${x}^{j}$ is the proper distance of 
the pendulum from the origin and ${R^j}_{0,k0}$ are the relevant components of the curvature tensor
in terms of the metric perturbation  $h_{\mu\nu}$ defined by\footnote{As is usual, latin indices run from $1-3$. Also $;$ denotes covariant derivatives.}   
\begin{eqnarray}
g_{\mu\nu} = \eta_{\mu\nu} + h_{\mu\nu}; \, |h_{\mu\nu}|<<1
\label{metric_perturbation}
\end{eqnarray}
on the flat Minkowski background $\eta_{\mu\nu}$.

The 
gauge-choice
$\left( h_{0\mu} = 0,  h_{\mu\nu;}{}^{\mu} =0, h_\mu^\mu=0 \right)$
that removes the unphysical degrees of freedom (DOF) and renders the GW transverse and traceless, has been made, so that only non-trivial components of the curvature tensor 
${R^j}_{0,k0} = -\ddot{h}_{jk}/2$
appear in eq.(\ref{e5}). The two physical DOF brought out by this choice are referred as the $\times$ and $+$ polarizations of GW and considering $z$ to be the propagation direction, the surviving components of the $2\times 2$ matrix $h_{jk}$ in the transverse 
plane, $h_{11} = -h_{22}$ and $ h_{12} = h_{21}$ represent these states respectively.

Note that, eq.(\ref{e5}) can be used as long as the spacial velocities involved are non-relativistic and $|{x}^{j}|$ is much smaller than 
the reduced wavelength $\frac{\lambda}{2\pi}$ of GW. These conditions are collectively referred as the \textit{small-velocity and long wavelength limit} and met by resonant bar-detectors and earth bound interferometric detectors\footnote{Note that these conditions are not satisfied by the proposed space-borne interferometer LISA  or by the Doppler tracking of spacecraft.} with the origin of the coordinate system centered at the detector. This also ensures that in a plane-wave expansion of GW, 
$h_{jk} = \int  (A_{jk} e^{ikx} + A^{*}_{jk} e^{- ikx})
d^{3}k/\left(2 \pi\right)^{3},$
the spatial part $e^{i \vec{k}.\vec{x}} \approx 1$ all over the detector site. Thus our only concern is the time-dependent part of the GW. If the polarization information contained in $A_{jk}$ is expressed in terms of the Pauli spin matrices, $h_{jk}$ takes the most convenient form 
\begin{equation}
h_{jk} \left(t\right) = 2f \left(\varepsilon_{\times}\sigma^1_{jk} + \varepsilon_{+}\sigma^3_{jk}\right) 
\label{e13}
\end{equation}
where $2f$ is the amplitude of the GW and $\left( \varepsilon_{\times}, \varepsilon_{+} \right)$ are the two possible polarization states of the GW satisfying the condition $\varepsilon_{\times}^2+\varepsilon_{+}^2 = 1$ for all $t$. For linearly polarized  GW the frequency $\Omega$ is contained in the time-dependent amplitude $2f(t)$ whereas for circularly polarized GW the time-dependent polarization states $\left( \varepsilon_{\times} \left(t \right), \varepsilon_{+} \left( t \right) \right)$ contains the frequency $\Omega$.

The Lagrangian for the system (\ref{e5}),  can be written, upto a total derivative term as 
\begin{equation}
{\cal L} = \frac{1}{2} m\dot {x}^2 - m{\Gamma^j}_{0k}
\dot {x}_{j} {x}^{k}  - \frac{1}{2} m \varpi^{2} \left(x_{j}\right)^2 
\label{e8}
\end{equation}
where ${R^j}_{0,k0} = - \frac{d \Gamma^j_{0k}}{d t}  = -\ddot{h}_{jk}/2  $.
Computing the canonical momentum ${p}_{j} = m\dot {x}_{j} - m \Gamma^j_{0k} {x}^{k}$  corresponding to ${x}_{j}$ we write the Hamitonian 
\begin{equation}
{H} = \frac{1}{2m}\left({p}_{j} + m \Gamma^j_{0k} {x}^{k}\right)^2 + \frac{1}{2} m \varpi^{2} \left(x_{j}\right)^2  \>.
\label{e9}
\end{equation}
Once we have the classical Hamiltonian we can have the NCQM description of the system simply by elevating the phase-space variables $\left( x^{j}, p_{j} \right)$ to operators $\left( {\hat x}^{j}, {\hat p}_{j} \right)$ and imposing the NC Heisenberg algebra (\ref{e9a}). 
Note that this algebra can be mapped \cite{cst, stern} to the standard $\left( \theta = 0 \right)$ Heisenberg algebra spanned by the operators $X_{i}$ and $P_{j}$ of the ordinary QM  through 
\begin{eqnarray}
{\hat x}_{i} = X_{i} - \frac{1}{2 \hbar} 
\theta \epsilon_{ij} P_{j}\>, \quad {\hat p}_{i} = P_{i} \>.
\label{e9b}
\end{eqnarray}
Writing the NCQM version of eq.$(\ref{e9})$ and employing the map (\ref{e9b}) 
we obtain\footnote{The traceless property of the GW is also required here.} 
\begin{eqnarray}
{\hat H} &=&\frac{ P_{j}{}^{2}}{2m} + \frac{1}{2} m \varpi^{2} X_{j}{}^{2}+ \Gamma^j_{0k} X_{j} P_{k} -\frac{m \varpi^{2}}{2 \hbar} \theta \epsilon_{jm} X^{j} P_{m} \nonumber \\ && 
-\frac{\theta }{2 \hbar} \epsilon_{jm} P_{m} P_{k}  \Gamma^j_{0k} \> = \hat{H}_{0} + \hat{H}_{{\rm {int}}}. 
\label{e12}
\end{eqnarray}
This Hamiltonian gives the commutative equivalent description of the noncommutative system (\ref{e9}) in terms of the operators $X_{i}$ and $P_{j}$. Since they admit the standard Heisenberg algebra, the rules of ordinary QM applies to (\ref{e12}). Also note that it has been demonstrated in various formulations of NC general relativity \cite{grav, grav1, banerjee11} that any NC correction in the gravity sector is second order in the NC parameter. Therefore, in a first order theory in NC space, the GW remains unaltered by NC effects.
The first two terms in eq.(\ref{e12}) represent the unperturbed HO Hamiltonian $\hat{H}_{0}$. Rest of the terms are small compared to $\hat{H}_{0}$ and can be treated as perturbations $\hat{H}_{{\rm {int}}}$. 
A term quadratic in $\Gamma$ has been neglected in eq.$(\ref{e12})$ since we deal with linearized gravity.

Defining raising and lowering operators in terms of the oscillator frequency $\varpi$
\begin{eqnarray}
X_j = \sqrt{ \frac{\hbar}{2m \varpi}}\left(a_j+a_j^\dagger\right); \quad
P_j = \sqrt{ \frac{\hbar m\varpi}{2i}} 
\left(a_j-a_j^\dagger\right)\,
\label{e15}
\end{eqnarray}
we write the time-dependent interaction part of the Hamiltonian (\ref{e12}) as 
\begin{eqnarray}
{\hat H}'_{{\rm {int}}}(t) &=& - \frac{i\hbar}{4} \dot h_{jk}(t) \left(a_j a_k - a_j^\dagger a_k^\dagger\right) \nonumber\\  
&&+ \frac{m \varpi \theta}{8} \epsilon_{jm} {\dot h}_{jk}(t)  \left(a_{m}a_{k}  - a_{m}a_{k}^\dagger + C.C \right) \nonumber\\ 
\label{e16}
\end{eqnarray}
where C.C means complex conjugate. 
We can now apply the time-dependent perturbation theory to compute the probability of transition between the ground state $|0,0\rangle$ and the  excited states of the 2-$d$ harmonic oscillator.
To the lowest order approximation 
the probability amplitude of transition from an initial state $|i\rangle$
to a final state $|f\rangle$, ($i\neq f$), due to a 
perturbation $\hat{V}(t)$ is given by \cite{kurt}
\begin{eqnarray}
C_{i \rightarrow f}(t\rightarrow\infty)  =  -\frac{i}{\hbar} \int_{-\infty}^{t\rightarrow +\infty}dt'
&& \left[ F_{jk}\left( t' \right)e^{\frac{i}{\hbar}(E_f -E_i)t'} \right.
\nonumber\\
&& \left. \times \langle \Phi_f |\hat{Q}_{jk}|\Phi_i \rangle \right]
\label{probamp}
\end{eqnarray}
where $\hat{V}(t)=F_{jk}(t)\hat{Q}_{jk}$.
Using the above result, we find that the {{\it{probability of transition  survives only between the ground state $|0,0\rangle$ and the second excited state}} and it reads
\begin{eqnarray}
C_{0\rightarrow 2} = - \frac{i}{\hbar} \int_{-\infty}^{+\infty} dt && \left[ F_{jk} \left( t \right) e^{\frac{i}{\hbar} \left( E_{2} - E_{0}\right)t} \left (\langle 2,0|\hat Q_{jk}|0,0\rangle \right. \right. \nonumber\\
&& \left. \left.+ \langle 1,1|\hat Q_{jk}|0,0\rangle + \langle 0,2|\hat Q_{jk}|0,0\rangle\right)\right] \nonumber\\
\label{trans_amp_02}
\end{eqnarray}
where $F_{jk} \left( t \right) = \dot h_{jk}(t)$ contains the explicit time dependence of ${\hat H}'_{{\rm {int}}}$ and 
\begin{eqnarray}
\hat Q_{jk} = && - \frac{i\hbar}{4} \left(a_j a_k - a_j^\dagger a_k^\dagger\right) \nonumber\\
&& + \frac{m \varpi \theta}{8} \epsilon_{jm}  \left( a_{m}a_{k}  - a_{m}a_{k}^\dagger + C.C \right) 
\label{Q}
\end{eqnarray}
contains the raising and lowering operators appearing in eq.(\ref{e16}). Expanding out ${\hat Q}$ for $i,j = 1,2$, we obtain the transition amplitude between the ground state $|0,0\rangle$ and the second excited state to be 
\begin{eqnarray}
C_{0\rightarrow 2} = - \frac{i}{\hbar} \int_{-\infty}^{+\infty} dt \, e^{2 i \varpi t} \left( \frac{i\hbar}{2} \dot h_{12}(t) + \frac{m \varpi \theta}{4} \dot h_{11}(t)\right).\nonumber\\
\label{trans_amp_02a}
\end{eqnarray}
The above equation is the main working formula in this paper.
Now using the general formula (\ref{trans_amp_02a}), we can compute the corresponding transition probabilities 
\begin{eqnarray}
P_{0\rightarrow 2} =  |C_{0\rightarrow 2}|^{2}
\label{trans_prob}
\end{eqnarray}
taking various template of gravitational wave-forms that are likely to be generated in runaway Astronomical events. 

We start with the simple scenario of periodic GW generically written as  
\begin{equation}
h_{jk} \left(t\right) = 2f_{0} \cos{\Omega t} \left(\varepsilon_{\times}\sigma^1_{jk} + \varepsilon_{+}\sigma^3_{jk}\right) 
\label{lin_pol}
\end{equation}
where the amplitude varies sinusoidally with a single frequency $\Omega$. In this limiting case of an exactly monochromatic wave the temporal duration of the signal is infinite and we get for the transition probability
\begin{eqnarray}
P_{0\rightarrow 2} &=&  \left( \pi f_{0} \Omega \right)^{2}  \left(\varepsilon_{\times}{}^{2}  + \Lambda^{2} \varepsilon_{+}{}^{2} \right) 
\nonumber \\ && \times 
\left[\delta \left(2 \varpi + \Omega \right) - \delta \left(2 \varpi - \Omega \right) \right]^{2}
\label{trans_prob_lin_pol}
\end{eqnarray}
where \begin{eqnarray}\Lambda = \frac{m \varpi \theta}{2 \hbar} = 1.888 \left( \frac{m}{10^{3}{\rm kg}}\right) \left( \frac{\omega}{1{\rm kHz}}\right)
\label{dim_less_NC}
\end{eqnarray}
is a dimensionless parameter carring the NC signature. Here we have used the stringent upper-bound\cite{carol} $|\theta| \approx 4 \times 10^{-40} {\rm m}^{2}$ for spatial noncommutativity and for reference mass and frequency used values appropriate for fundamental phonon modes of a bar detector \cite{ncgw_4} which are formally identical to the NC harmonic oscillator system considered here.
 
Consider periodic GW signal coming from a binary system (with quasi-circular orbit) being received by some earth-bound detector; if the orbit of the binary system is edge-on with respect to us, then we receive the $+$polarization of the radiation only \cite{Magg} , i.e., $\left(\varepsilon_{\times}, \varepsilon_{+} \right) = \left( 0, 1 \right)$, and in this case (\ref{trans_prob_lin_pol}) shows that the transition probability will scale quadratically with the dimensionless parameter $\Lambda$ characterized by spatial noncommutativity. Therefore such a transition will be driven by the combined perturbative effect of GW as well as spatial noncommutativity\footnote{This corresponds to the last term in ${\hat H}_{{\rm {int}}}$ in (\ref{e16}).} and will occur only if the space has a NC structure. In other words, {\it a quantum mechanical transition induced by the linearly polarized GW from a binary system with its orbital plane lying parallel to our line of sight
can be an effective test of the noncommutative structure of space.} Note that the angular frequency $\Omega$ of the quadrupole radiation is twice the angular frequency of rotation of the source \cite{Magg}, thus (\ref{trans_prob_lin_pol}) also tells us that {\it for transition to occur we need to have a harmonic oscillator with natural frequency $\varpi$ that matches with that of the source.} Highly accurate X-ray/radio-astronomical measurement of the frequency of orbital rotation of binary Pulsars can be used to pin-point the natural frequency of the harmonic oscillator required here. Since response of most operating bar detectors \cite{bar_2} at present are centred around $0.9$kHz, we can alternatively assess the target source-type by estimating the orbital radius of the source binaries which can emit GW of a suitable frequency to enable the said transition. Using Kepler's law to express the orbital radius in terms of the angular frequency of rotation (= natural frequency of the harmonic oscillator in the present case) and total mass of the binary system, we find $R = \left\{\frac{G \left( m_{1} + m_{2} \right)}{\varpi^{2}}\right\}^{\frac{1}{3}}  \approx 23 {\rm km}$ where $ m_{1} = m_{2} = 1.4 M_{\odot}$, which is the lower mass-limit for neutron stars in terms of the Solar mass $M_{\odot}$, have been used as reference. Such small orbital separation can only be reached by highly compact binaries consisting of neutron stars and black holes \footnote{Note that since the radius of a neutron star of mass $1.4 M_{\odot}$ is about $10$ km, the point-mass approximation and Newtonian gravity limit assumed here can only provide a crude estimate. See the discussion in \cite{Magg} chapter 4.1.1.}. 

For completeness let us mention that 
the amplitude $f_{0}$ in eq.(\ref{lin_pol}) for quadrupole radiation \footnote{The power carried by a GW has a null dipole term, because of the momentum conservation law. The first
non-trivial component cois the quadrupole one.} of GW from an inspiral binary with masses $m_{1}, m_{2}$ and quasi-circular orbit is
\begin{eqnarray}
f_{0} = \frac{1}{r}\left(\frac{G M_{c}}{c^{2}}\right)^{\frac{5}{3}}\left(\frac{\Omega}{2c}\right)^{\frac{2}{3}} 
\label{bin_amp}
\end{eqnarray}
where $M_{c} = \frac{\left(m_{1} m_{2}\right)^{\frac{3}{5}}}{\left(m_{1} + m_{2}\right)^{\frac{1}{5}}}$ is the chirp mass of the binary and $r$ is the distance of the binary from the Earth. 

From another binary system similar to the one considered above, but with its orbital plane perpendicular to our line of sight, both the $+$ and $\times$ polarization of the radiation will reach the detector with equal amplitude and consequently we will have a source for circularly polarized GW signal that can be generically written as
\begin{equation}
h_{jk} \left( t \right) = 2f_{0} \left[\varepsilon_{\times} \left( t \right) \sigma^1_{jk} + \varepsilon_{+}\left( t \right) \sigma^3_{jk}\right] 
\label{cir_pol}
\end{equation}
with $\varepsilon_{+} \left( t \right)  = \cos \Omega t $ and $\varepsilon_{\times} \left( t \right)  = \sin \Omega t $ and amplitude given by eq.(\ref{bin_amp}). The transition probability in this case is 
\begin{eqnarray}
P_{0\rightarrow 2} & =& \left( \pi f_{0} \Omega \right)^{2}  \left[ \left\{ \left( 1 + \Lambda \right) \delta \left( 2 \omega + \Omega \right) \right\}^{2} \right.  \nonumber \\
&& \left.+ \left\{ \left( 1 - \Lambda \right) \delta \left( 2 \omega - \Omega \right) \right\}^{2} \right].
\label{trans_prob_cir_pol}
\end{eqnarray}
First let us note that as in the previous case, the transition occurs when natural frequency of the harmonic oscillator resonates with that of the source of the GW. Also notable is that here the transition probability has terms both linear and quadratic in the dimensionless NC parameter $\Lambda$. However eq.(\ref{dim_less_NC}) shows that for phonon modes in a bar detector which are the realization of the NC HO system considered in this paper, $\Lambda$ is of the order of unity\footnote{at least with the current upper-bound on the NC parameter $\theta$ \cite{carol}.}, so we cannot drop the quadratic term even though we started with a theory to first order in the NC parameter. Significantly, eq.(\ref{trans_prob_cir_pol}) shows a non-zero transition probability for $\Lambda = 0$, i.e. if our space has commutative structure. {\it Thus a transition induced by circularly polarized GW from a binary system cannot be used as a deterministic probe for spatial noncommutativity.} This feature lies with the earlier case of linearly polarized GW signals only.

In the last stable orbit of an inspiraling neutron star or black hole binary or during its merging and final ringdown, the system can liberate large amount of energy in GWs within a very short duration $10^{-3} {\rm sec} < \tau_{{\rm g}} < 1 {\rm sec}$. Such signals are referred to as GW bursts. Supernova explosions and stellar gravitational collapse are other candidate generators. Since bursts originate from violent and explosive astrophysical phenomena, their waveform cannot be accurately predicted and only be crudely modelled as
\begin{eqnarray}
h_{jk} \left(t\right) = 2f_{0} g \left( t \right) \left(\varepsilon_{\times}\sigma^1_{jk} + \varepsilon_{+}\sigma^3_{jk}\right) 
\label{lin_pol_burst}
\end{eqnarray}
where to be generic we have kept both components of the linear polarization. Here $g \left( t \right)$ is a smooth function which goes to zero rather fast for $|t| > \tau_{{\rm g}} $. A convenient choice is a function peaked at $t = 0$ with $g \left( 0 \right) = \mathcal{O}(1)$ so that $|h_{jk} \left( t \right) | \sim \mathcal{O}\left( f_{0} \right) $ near the peak. So we take a simple Gaussian 
\begin{equation}
g \left(t\right) = e^{- t^{2}/ \tau_{g}^{2}}.
\label{burst_waveform_Gaussian}
\end{equation}
Owing to its small temporal duration the burst have a continuum spectrum of frequency over a broad range upto $f_{{\rm max}} \sim 1/ \tau_{{\rm g}}$ whereas the detector is sensitive only to a certain frequency window and blind beyond it. If the sensitive band-width is small compared to the typical variation scale of the signal in the frequency space, the crude choice in eq.(\ref{burst_waveform_Gaussian}) instead of a precise waveform is good enough. In terms of the Fourier decomposed modes the GW burst can thus be modelled as 
\begin{eqnarray}
h_{jk} \left(t\right) = \frac{f_{0}}{\pi} \left(\varepsilon_{\times}\sigma^1_{jk} + \varepsilon_{+}\sigma^3_{jk}\right)  \int_{-\infty}^{+\infty} \tilde{g} \left( \Omega \right) e^{- i \Omega t}  d \Omega 
\label{lin_pol_burst_Gaussian}
\end{eqnarray}
where $\tilde{g} \left( \Omega \right) = \sqrt{\pi} \tau_{g} e^{- \left( \frac{\Omega \tau_{g}}{ 2} \right)^{2}}$ is the amplitude of the Fourier mode at frequency $\Omega$.

Using eq.(\ref{lin_pol_burst_Gaussian}) in the general formula for transition amplitude (\ref{trans_amp_02a}) and further employing eq.(\ref{trans_prob}), we find the probability for transition from the ground state to the second excited state induced by a GW burst is
\begin{eqnarray}
P_{0\rightarrow 2} & = &\left( 2 \sqrt{\pi}  f_{0} \varpi \tau_{g} \right)^{2} e^{- 2 \varpi^{2} \tau_{g}^{2}} \left( \varepsilon_{\times}{}^{2}  + \Lambda^{2} \varepsilon_{+}{}^{2} \right)
\label{trans_prob_Gaussian_burst}
\end{eqnarray}
where the GW Fourier mode with twice the natural frequency of the harmonic oscillator (the detector in our consideration) gets picked up. Since the burst signal duration $\tau_{g} \sim 10^{- 2} - 10^{- 3} {\rm sec} $, the maximum frequency in the Fourier spectrum can be $\Omega_{{\rm max}} /2\pi \sim 0.1 - 1 {\rm kHz}$ which partially overlaps with the sensitive bandpass for the bar-detectors. Actually the lower limit is in the bandpass of LIGO which, though, does not represent our harmonic oscillator system as closely as a bar-detector, still works in the long wavelength-low velocity limit considered here. 

From eq.(\ref{trans_prob_Gaussian_burst}) we again see that the $+$ polarization of the GW burst can only induce a transition if the space has a NC structure. As mentioned earlier the polarization state of a GW signal from a given source can be anticipated since it depends largely on the orientation of the source which can be determined by observing its electromagnetic radiation and the detector geometry which is at our disposal. {\it So detecting a QM transition induced by a GW burst from an appropriate source can serve as a probe of the spatial noncommutativity.} 

A slightly more realistic waveform can be realized if the Gaussian function in eq.(\ref{burst_waveform_Gaussian}) is modulated by some frequency $\Omega_{0}/2 \pi$, resulting in a sine-Gaussian amplitude
\begin{equation}
g \left(t\right) = e^{- t^{2}/ \tau_{g}^{2}}  \,  \sin \Omega_{0}t~.
\label{burst_waveform_sine_Gaussian}
\end{equation}
Then the signal in eq.(\ref{lin_pol_burst_Gaussian}) admits the Fourier modes
\begin{eqnarray}
\tilde{g} \left( \Omega \right) = \frac{i \sqrt{\pi} f_{0} \tau_{g}}{2} \left[ e^{- \left(\Omega - \Omega_{0}\right)^{2}\tau_{g}^{2}/4} - e^{- \left(\Omega + \Omega_{0}\right)^{2}\tau_{g}^{2}/4} \right]
\label{sine-Gaussian_Fourier}
\end{eqnarray}
centered around $\Omega =  \Omega_{0}$ in the frequency space.
The corresponding transition probability is 
\begin{eqnarray}
P_{0\rightarrow 2} & = & \left[ e^{- \left(2 \varpi - \Omega_{0}\right)^{2}\tau_{g}^{2}/4} - e^{- \left(2 \varpi + \Omega_{0}\right)^{2}\tau_{g}^{2}/4} \right]^{2} \nonumber\\
&& \times  \left( \sqrt{\pi}  f_{0} \varpi \tau_{g} \right)^{2} \left( \varepsilon_{\times}{}^{2}  + \Lambda^{2} \varepsilon_{+}{}^{2} \right)
\label{trans_prob_sine_Gaussian_burst}
\end{eqnarray}
where, as usual, only the appropriate Fourier mode is picked up by the harmonic oscillator. Two limiting cases can be immediately identified in eq.(\ref{trans_prob_sine_Gaussian_burst}). If the operating frequency of the detector is low, say, in the sub-Hz bandpass, the two exponential terms are of same size and nearly cancel each other, reducing the transition probability. The other extreme is when $2 \varpi - \Omega_{0} = \Delta \varpi$ with $\frac{\Delta \varpi}{\varpi} << 1$. Then the first exponential term is sizeable while the second term is negligible in comparison and we get a transition amplitude 
\begin{eqnarray}
P_{0\rightarrow 2}  \approx   e^{- \left(\Delta \varpi \right)^{2}\tau_{g}^{2}/2}  \left( \sqrt{\pi}  f_{0} \varpi \tau_{g} \right)^{2} \left( \varepsilon_{\times}{}^{2}  + \Lambda^{2} \varepsilon_{+}{}^{2} \right).
\label{trans_prob_sine_Gaussian_burst_1}
\end{eqnarray} 
Note that since $\tau_{g} \sim2 \pi/\Omega_{\rm max} $, the quantity $\left( \Delta \varpi \tau_{g} \right)^{2} \sim \left(2 \pi \Delta \varpi /\Omega_{\rm max}\right)^{2}$. For bar detectors $\Delta \varpi$ is at least within the sensitive bandpass \cite{bar_1}, so it cannot be more than a few Hz, whereas $\Omega_{\rm max}$ is in the kHz range for GW burst of a few millisecond duration. This ensures that the Gaussian factor in eq.(\ref{trans_prob_sine_Gaussian_burst_1}) suppresses the transition probability only marginally. Interestingly, the Gaussian factor in eq.(\ref{trans_prob_Gaussian_burst}) is much stronger compared to that in eq.(\ref{trans_prob_sine_Gaussian_burst_1}), since $\varpi \tau_{g} >>  \Delta \varpi \tau_{g}$. Thus for the more realistic sine-Gaussian template, the transition probability is actually higher. 

In conclusion we would like to convey that the considerations in the present paper suggest that the joint operation of various resonant detector groups like ALLEGRO, AURIGA, EXPLORER, NAUTILUS and NIOBE around the world in IGEC (InternationalGravitational Event Collaboration) \cite{bar_detectors_1, bar_detectors_2, bar_detectors_3, bar_detectors_4, bar_detectors_5}  may possess the potential to establish the possible existance of a granular structure of our space as a by-product in the event of a direct detection of GW and therefore must be continued. 

\section*{Acknowledgment} \noindent AS and SG thank IUCAA where the work was done. AS acknowledges the finantial support of DST SERB under Grant No. SR/FTP/PS-208/2012. 


\end{document}